\documentclass[amsmath,amssymb,12pt]{iopart}

\usepackage{amssymb}
\usepackage{graphicx}
\usepackage{dcolumn}
\usepackage{bm}
\usepackage{verbatim}
\usepackage{wrapfig}
\usepackage{ascmac}
\usepackage{makeidx}
\usepackage{mathrsfs}
\usepackage{braket}
\usepackage{cite}
\newcommand{\MnTaO}{Mn$_{3}$Ta$_{2}$O$_{8}$}

\begin{document}

\title{Spin dynamics in linear magnetoelectric material \MnTaO\ }

\author{Hodaka Kikuchi$^1$, Shunsuke Hasegawa$^1$, Shinichiro Asai$^1$, Tao Hong$^2$, Kenta Kimura$^3$, Tsuyoshi Kimura$^4$, Shinichi Itoh$^5$, Takatsugu Masuda$^{1,6,7}$}

\address{$^1$ Institute for Solid State Physics, University of Tokyo, Kashiwa, Chiba 277-8581, Japan}
\address{$^2$ Neutron Scattering Science Division, Oak Ridge National Laboratory, Oak Ridge, Tennessee 37831-6393, USA}
\address{$^3$ Department of Materials Science, Osaka Metropolitan University, Sakai, Osaka 599-8531, Japan}
\address{$^4$ Department of Applied Physics, University of Tokyo, Bunkyo-ku, Tokyo 113-8656, Japan}
\address{$^5$ Neutron Science Division, Institute of Materials Structure Science, High Energy Accelerator Research Organization, Tsukuba, Ibaraki 305-0801, Japan}
\address{$^6$ Institute of Materials Structure Science, High Energy Accelerator Research Organization, Ibaraki 305-0801, Japan}
\address{$^7$ Trans-scale Quantum Science Institute, The University of Tokyo, Tokyo 113-0033, Japan}
\ead{hodaka.kikuchi@issp.u-tokyo.ac.jp}
\vspace{10pt}
\begin{indented}
\item[]Apr 2024
\end{indented}

\begin{abstract}
We performed inelastic neutron scattering experiments on single crystal samples of a linear magnetoelectric material \MnTaO , which exhibits a collinear antiferromagnetic order, 
to reveal the spin dynamics. 
Numerous modes observed in the neutron spectra were reasonably reproduced by linear spin-wave theory on the basis of the spin Hamiltonian including eight Heisenberg interactions and an easy-plane type single-ion anisotropy.
The presence of strong frustration was found in the identified spin Hamiltonian.
\end{abstract}

%
\vspace{2pc}

\section{Introduction}

\begin{figure}
\begin{center}
\includegraphics[width=8cm]{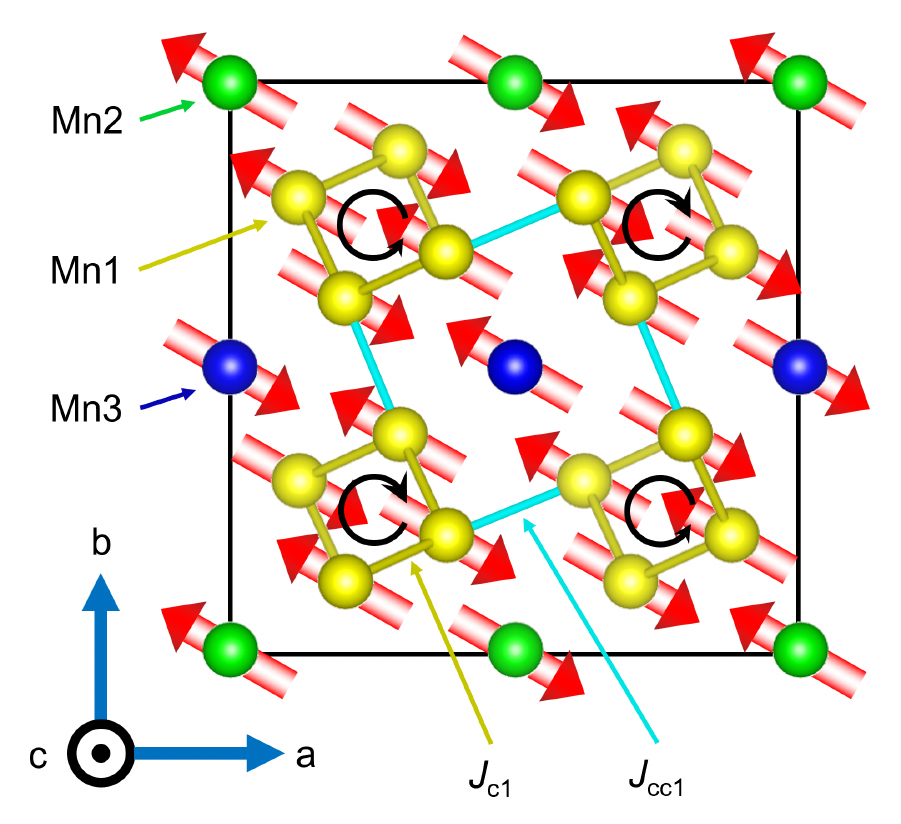}
\caption{The crystal and magnetic structures of \MnTaO\ projected onto the crystallographic $ab$ plane (omitting Ta and O). Yellow circles represent Mn$^{2+}$ ions at the Mn1 site, green ones represent Mn$^{2+}$ ions at the Mn2 site, and blue ones represent Mn$^{2+}$ ions at the Mn3 site. Mn$^{2+}$ ions at the Mn1 site form a spiral structure along the $c$-axis, with arrows indicating the chirality of the spiral. Red arrows represent the magnetic moments. $J_{\rm c1}$, indicated by the yellow bond, is antiferromagnetic, and $J_{\rm cc1}$, indicated by the light blue bond, is ferromagnetic; they compete with each other.}
\label{structure}
\end{center}
\end{figure}

The interaction between magnetism and electricity induces a magnetoelectric (ME) effect, in which the electric dipole moment and magnetic moment are simultaneously induced. 
 This phenomenon was first observed in Cr$_{2}$O$_{3}$\cite{ASTROV1960,Rado61}, characterized by an electric polarization and a magnetic moment that respond linearly to applied magnetic and electric fields, respectively. 
To exhibit the linear ME effect, it is necessary to break both spatial inversion and time-reversal symmetries. 
Since the discovery in Cr$_{2}$O$_{3}$, the effect has been studied in various materials\cite{Fiebig2005,Kornev2000,Mufti2011,Jiang2018}, 
becoming one of the most active research topics in the field of condensed matter physics in recent years.

In frustrated magnets exhibiting noncollinear orders such as cycloidal structures\cite{Kimura2003,Taniguchi2006,Sergienko2006,Kenzelmann2005,Arkenbout2006}, spontaneous electric polarization and ordered magnetic moments are simultaneously induced via the spin-current mechanism\cite{Katsura2005}. Other contributing mechanisms include exchange striction\cite{Arima2006,Okuyama2011} and spin-dependent metal-ligand hybridization\cite{Arima2007,Murakawa2010,soda2014}.
In $R$MnO$_3$ ($R$ = rare earth), numerous inelastic neutron scattering (INS) experiments have been conducted, revealing the presence of the frustration~\cite{Kajimoto2005,Joosung2016,Holbein2023}.

\MnTaO\ is a tetragonal crystal with space group $I4_{1}/a$, where Mn$^{2+}$ ions 
carry spin $S$ = $5/2$~\cite{Esmaeilzadeh1998,Grins1999, Rickert2015,Kimura2021}.
Figure \ref{structure} illustrates the crystal and magnetic structures, 
projecting the Mn$^{2+}$ ions onto the $ab$ plane, with different Mn sites indicated by different colors. 
Mn$^{2+}$ ions at Mn1 sites represented by yellow symbols form spirals along the $c$-axis, with the black arrows indicating the chiralities of the spirals.
At $T_{\rm N}$ = 24 K, a magnetic phase transition to a collinear antiferromagnetic structure occurs.
The moment sizes of Mn$^{2+}$ ions at Mn1, Mn2, and Mn3 sites are 3.91(3), 4.17(8) and 4.04(8) $\mu_{\rm B}$, respectively\cite{Kimura2021}. 
Previous studies \cite{Grins1999,Kimura2021} have revealed that the Weiss temperature($\theta _{\rm W}$) is $-$120 K, 
suggesting strong frustration, as $\theta _{\rm W}$ is five times larger than $T_{\rm N}$.
The magnetic structure belongs to the $2'/m$ space group, breaking both spatial inversion and time-reversal symmetries, leading to the observation of linear magnetoelectric effect. 
An antiferroelectric-like phase transition due to the off-center displacement of magnetic Mn$^{2+}$ ions has been observed \cite{Kimura2021}.

In the present study, INS experiments were performed using single crystal samples to elucidate the spin dynamics in \MnTaO. 
The observed spectra were reasonably explained by the spin Hamiltonian composed of eight Heisenberg terms and an easy-plane single-ion anisotropy term. 
Strong frustration was found between the intrachain antiferromagnetic interaction and interchain ferromagnetic interaction.

\section{Experimental details}

Single crystals of \MnTaO\ were grown by the floating-zone method\cite{Kimura2021}. 
The quality of the samples was evaluated  using High Energy Resolution (HER) triple-axis spectrometer installed at the C11 beam port at Japan Research Reactor 3 (JRR-3).

To collect INS spectra in wide momentum ($\bm{q}$) - energy ($\hbar \omega$) space, INS experiments were carried out by High Resolution Chopper spectrometer (HRC)~\cite{ITOH201190} installed at BL12 in J-PARC/MLF. 
A single crystal of \MnTaO\ with a mass of 2.20 g was used. 
Temperatures were controlled using a Gifford-McMahon (GM) cryostat. 
Frequency of Fermi chopper was 100 Hz, and incident neutron energy ($E_{\rm i}$) was 15 meV.
The instrumental energy resolution at $\hbar\omega$ = 0 is 0.49 meV.
The data reduction was performed by HANA software~\cite{Kawana_2018}.

To detail the spectra in low-energy range, INS experiments were carried out by cold neutron triple-axis spectrometer (CTAX) installed at High Flux Isotope Reactor (HFIR) in Oak Ridge National Laboratory (ORNL).
A single crystal with a mass of 1.57 g was used.
Cooling was achieved with an ORANGE cryostat utilizing liquid helium, maintaining the measurement temperature at 1.5 K. 
The collimator setup was configured as guide - open - radial collimator - 120'.
Final neutron energy was fixed at $E_{\rm f} = 3.5$ meV.
BeO filter was employed downstream of the analyzer to cut off the higher harmonics.
The instrumental energy resolution ranges from 0.22 meV to 0.40 meV, depending on the energy transfer.

\section{Experimental results}

\begin{figure*}
\begin{center}
\includegraphics[width=15.5cm]{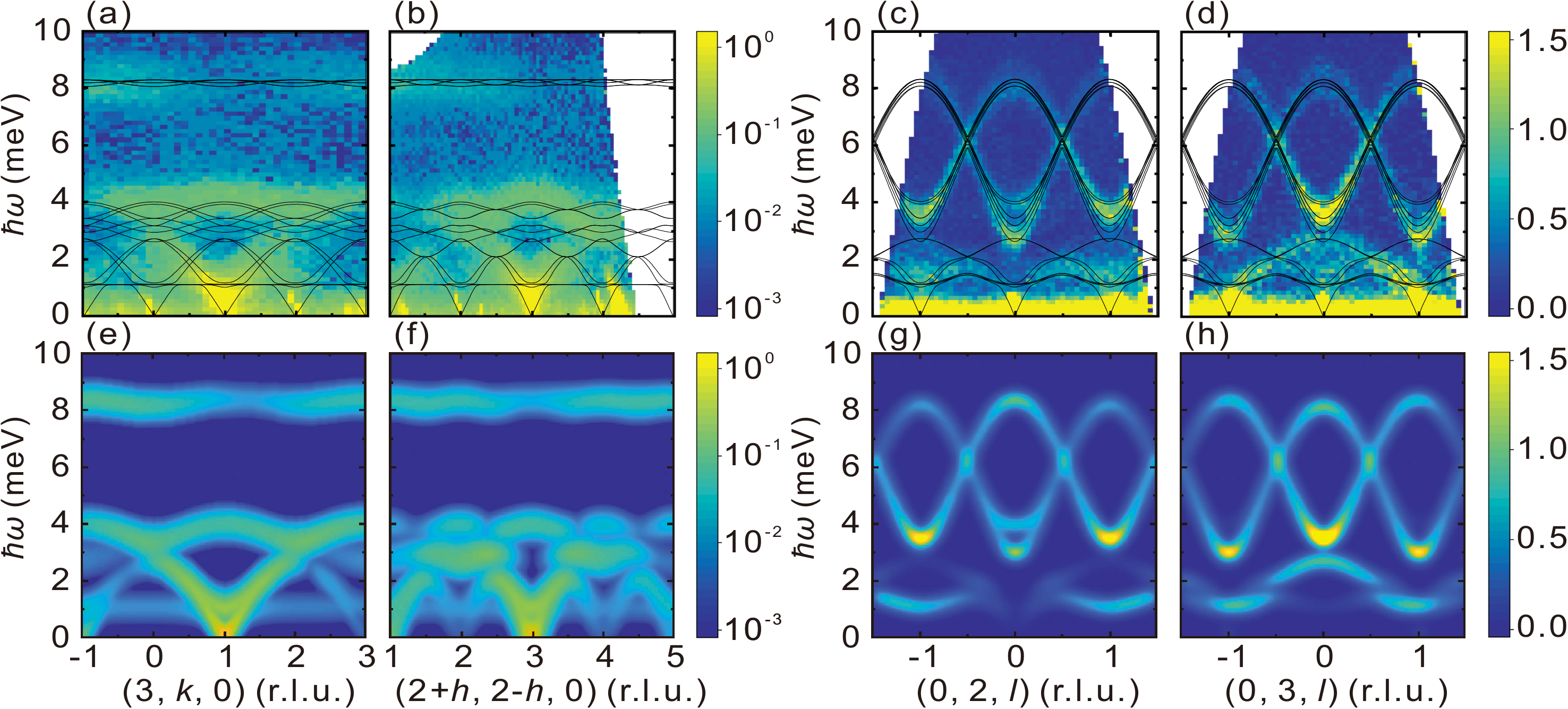}
\caption{(a)-(d) Inelastic neutron scattering (INS) spectra measured using the HRC spectrometer. Solid curves show the dispersion curves calculated by linear spin wave theory (LSWT). (e)-(h) INS spectra calculated by LSWT. }
\label{HRC_exp}
\end{center}
\end{figure*}

\begin{figure}
\begin{center}
\includegraphics[width=14cm]{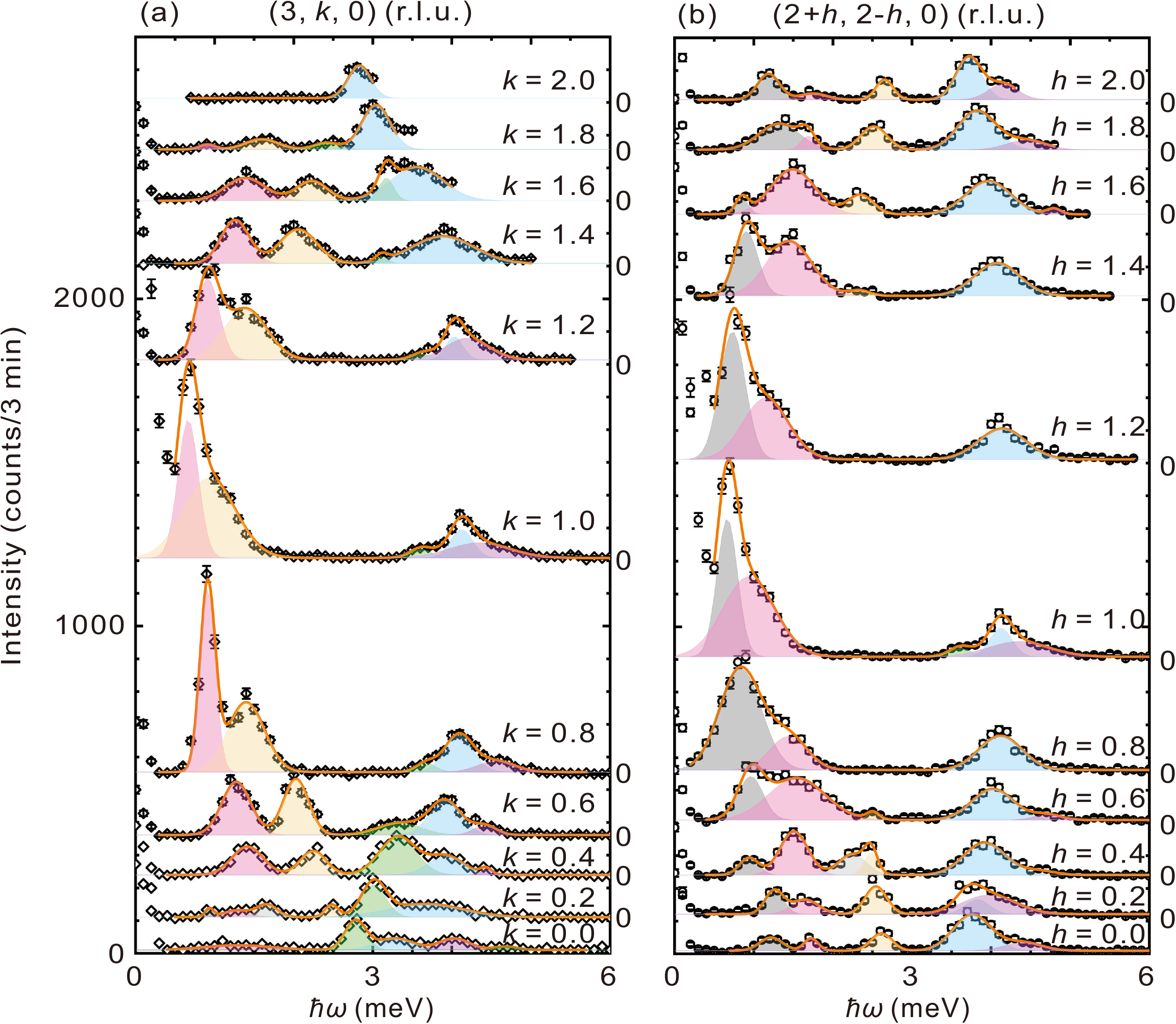}
\caption{INS spectra measured using CTAX spectrometer along (a) $(3,\ k,\ 0)$ and (b) $(2+h,\ 2-h,\ 0)$. The symbols represent the experimental data. The orange curves are the fitted results using multiple Gaussian functions.}
\label{CTAX_exp}
\end{center}
\end{figure}

First, we conducted measurements over a wide range of $\bm{q}$-$\hbar\omega$ space using the HRC spectrometer. 
In Figs \ref{HRC_exp}(a) and \ref{HRC_exp}(b), gapless excitations were observed, arising from $(3,\ 1,\ 0)$ in the range of $\hbar\omega\  \lesssim$ 3 meV.
Other excitations were observed in the range of 3 meV $\lesssim\ \hbar\omega\ \lesssim$ 5 meV, and flat excitations were observed near 8 meV.
Note that the scale of the intensity is logarithmic. 
In Figs. \ref {HRC_exp}(c) and \ref{HRC_exp}(d), low energy excitations are observed in the range of $\hbar \omega \lesssim 3$ meV.
Further observations showed modes in the range of 3 meV $\lesssim\ \hbar\omega\ \lesssim$ 8 meV. 
The solid curves in Figs. \ref{HRC_exp}(a)-\ref{HRC_exp}(d) represent the dispersion relations in each direction, calculated using the linear spin-wave theory (LSWT), the details of which will be described later.

Next, we conducted measurements using CTAX, which provides a higher energy resolution, to elucidate the details of low-energy excitations. 
The constant $\bm{q}$ scans were performed centered at (3,\ 1,\ 0), along $(0,\ k,\ 0)$ and $(h,\ -h,\ 0)$ as shown in Figs \ref{CTAX_exp}(a) and \ref{CTAX_exp}(b), respectively. 
The results are consistent with those measured by HRC spectrometer in Figs. \ref{HRC_exp}(a) and \ref{HRC_exp}(b).
Solid curves represent the fitting curves obtained through multiple Gaussian fittings. The full widths at half maximum (FWHM) of most of the Gaussians are broader than the instrumental resolution. 
Due to the large number of modes and some having weak intensities, it is not feasible to separate all of them for fitting.

\section{Analysis}

\begin{figure}
\begin{center}
\includegraphics[width=10cm]{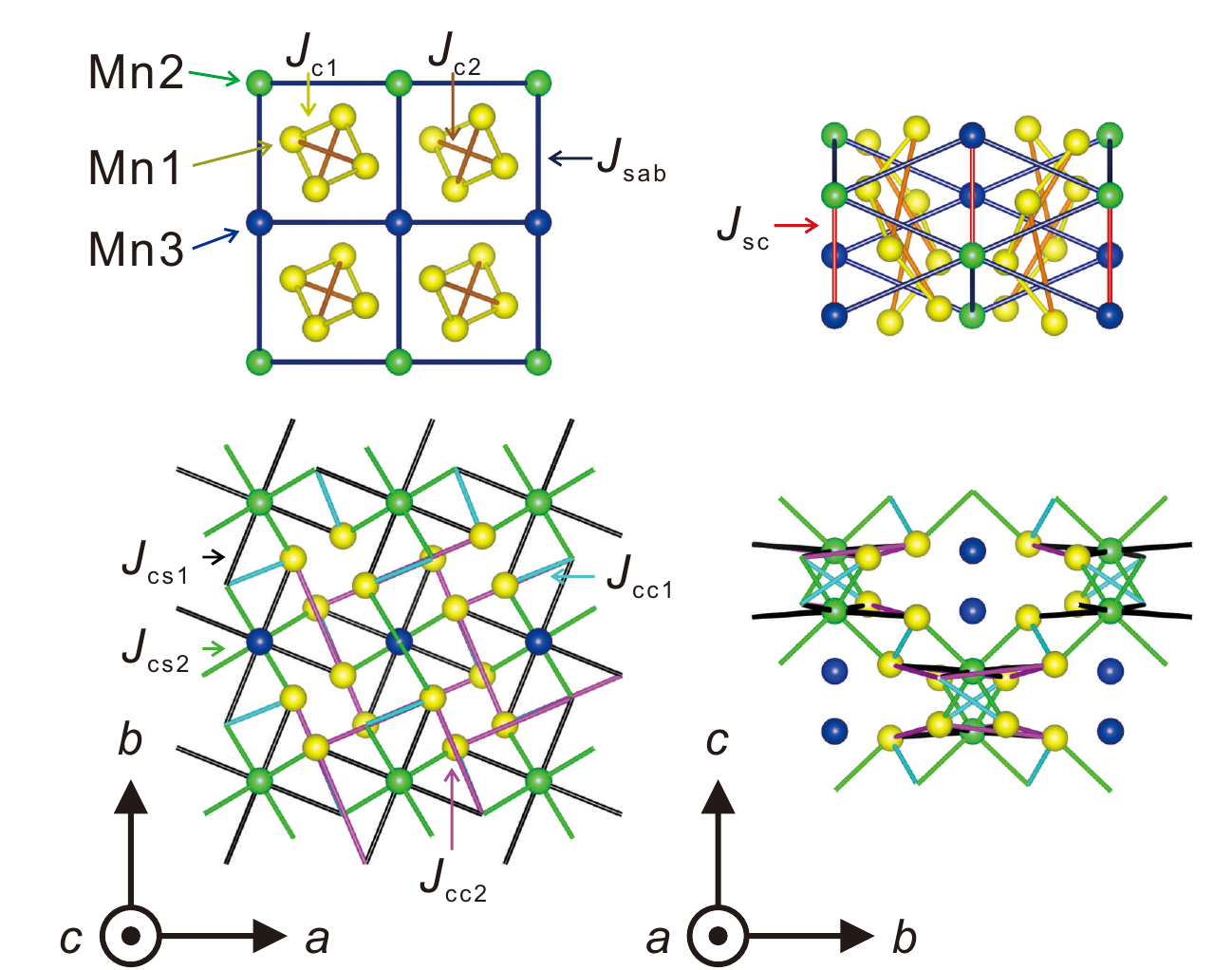}
\caption{Interaction pathways in \MnTaO. The intrachain and intrasquare interactions projected onto the $ab$ and $bc$ planes are illustrated in (a) and (b), respectively. The chain is formed by the Mn$^{2+}$ ions at the Mn1 sites and the square is formed by those at the Mn2 and Mn3 sites. The interactions between the chain and square projected onto the $ab$ and $bc$ planes are illustrated in (c) and (d), respectively.}
\label{ana1}
\end{center}
\end{figure}

The observed INS spectra were analyzed based on LSWT by using the SpinW package\cite{spinW}. The spin Hamiltonian is as follows; 
\begin{eqnarray}
{\mathcal H}&=&J_{\rm c1}\sum_{i.j}\bm{S}_{i}\cdot\bm{S}_{j}+J_{\rm c2}\sum_{i,k}\bm{S}_{i}\cdot\bm{S}_{k}+J_{\rm sab}\sum_{i,l}\bm{S}_{i}\cdot\bm{S}_{l} \nonumber \\
&+&J_{\rm sc}\sum_{i,m}\bm{S}_{i}\cdot\bm{S}_{m}+J_{\rm cc1}\sum_{i,n}\bm{S}_{i}\cdot\bm{S}_{n}+J_{\rm cc2}\sum_{i,o}\bm{S}_{i}\cdot\bm{S}_{o} \nonumber \\
&+&J_{\rm cs1}\sum_{i,p}\bm{S}_{i}\cdot\bm{S}_{p}+J_{\rm cs2}\sum_{i,q}\bm{S}_{i}\cdot\bm{S}_{q}+D\sum_{i}^{\rm site}(\bm{S}_{z})^{2}.
\label{eq1}
\end{eqnarray}
The magnetic structure used for the analysis is depicted in Fig. \ref{structure}. Even though Mn$^{2+}$ ions at Mn1, Mn2, and Mn3 sites have different moment sizes, 
we assumed a uniform moment size of 4 $\mu_{\rm B}$ for all Mn$^{2+}$ ions. 
Each interaction is defined in Fig. \ref{ana1}. 
For convenience, we designated the structure composed of Mn$^{2+}$ ions at Mn1 site as ``chain'' and the structure composed of Mn$^{2+}$ ions at Mn2 and Mn3 sites as ``square''. 
Note that the projection onto the $ab$ plane appears ``square'' as shown in Fig.~\ref{ana1}(a); however, the structure is not a true square as depicted in Fig.\ref{ana1}(b). 
In Figs \ref{ana1}(c) and \ref{ana1}(d) the interactions between chain and square are projected onto the $ab$ and $bc$ planes, respectively. 
In addition to the eight interactions shown here, an easy-plane type single-ion anisotropy was incorporated into the Hamiltonian.

LSWT analysis was performed on the dispersion relation measured along $(3,\ k,\ 0)$ and $(2+h,\ 2-h,\ 0)$ by the CTAX spectrometer and that along $(0,\ 2,\ l)$ and $(0,\ 3,\ l)$ by the HRC spectrometer. The weighted sum of squared residuals is calculated as follows: 
\begin{eqnarray}
\chi^2=\frac{1}{N}\sum_{i,j}\frac{(\hbar\omega_{i}^{\rm{exp}}(\bm{q}_{j})-\hbar\omega_{i}^{\rm{cal}}(\bm{q}_{j}))^2}{\sigma_{i}^2}.
\label{eq2}
\end{eqnarray}
Here, $\hbar\omega_{i}^{\rm exp}(\bm{q}_{j})$ is the observed energy of the $i$-th peak in a constant $q$-scan/slice at $\bm{q}$ = $\bm{q}_j$, $\hbar\omega_{i}^{\rm cal}(\bm{q}_{j})$ is the calculated excitation energy that is the closest to $\hbar\omega_{i}^{\rm exp}(\bm{q}_{j})$,  $\sigma_i$ represents the FWHM of the Gaussian fit to the excitation at the $i$-th data point, and $N$ is the number of data points. 
Given that multiple modes are in close proximity to one another, the FWHM was used as the weighting factor in the fitting process for parameter optimization in the analysis.

\begin{table}
\caption{\label{tab1}The parameters of the spin Hamiltonian estimated by linear spin wave theory.} 

\begin{indented}
\lineup
\item[]\begin{tabular}{ccccc}
$J_{\rm c1}$ (meV) & $J_{\rm c2}$ (meV) & $J_{\rm sab}$ (meV) & $J_{\rm sc}$ (meV)\\
$1.35(5)$ & $-0.22(2)$ & $0.05(1)$ & $-0.20(2)$\\
\hline
$J_{\rm cc1}$ (meV) & $J_{\rm cc2}$ (meV) & $J_{\rm cs1}$ (meV) & $J_{\rm cs2}$ (meV) & $D$ (meV)\\
$-0.22(2)$ & $-0.03(1)$ & $0.26(2)$ & $-0.05(1)$ & $0.04(1)$\\
\end{tabular}
\end{indented}
\end{table}

The $\chi ^2$ values were computed for all nine parameters varied within a wide range, and the parameters with the lowest $\chi^2$ values are summarized in Table \ref{tab1}. 
The values in parentheses represent the step size of each parameter used during the fitting. 
The dispersion curves calculated using the values from Table \ref{tab1} are shown in Figs. \ref{HRC_exp}(a)-\ref{HRC_exp}(d), and the INS spectra are shown in Figs. \ref{HRC_exp}(e)-\ref{HRC_exp}(h). 
The calculations reasonably reproduce the experimental spectra, and we identified the spin Hamiltonian.

\section{Discussion}

In Fig. \ref{structure}, $J_{\rm c1}$, indicated by a yellow bond, is antiferromagnetic, and $J_{\rm cc1}$, indicated by a light blue bond, is ferromagnetic. The neighboring spins connected by $J_{\rm cc1}$ are antiparallel to each other, which demonstrates the presence of frustration between $J_{\rm c1}$ and $J_{\rm cc1}$.
Our calculations suggest that a noncollinear magnetic structure emerges when $J_{\rm cc1}$ exceeds 30\% of $J_{\rm c1}$. In the identified spin Hamiltonian, however, $J_{\rm cc1}$ is 16\% of $J_{\rm c1}$ and it is not strong enough to induce the noncollinear structure.

The curvatures of INS spectra near 4 meV around (3, 1, 0) in Figs. \ref{HRC_exp}(a), 2(b), 2(e), and 2(f) are negative. 
We confirmed that the curvature is, in fact, influenced by the presence or absence of frustration.
In cases of positive $J_{\rm cc1}$ --indicative of no frustration in the magnetic structure depicted in Fig.~\ref{structure}--
the calculated curvature was positive, contrasting with the experimental observations. 
The presence of frustration, thus, qualitatively affects the spectra.

The magnitude of the ME tensor, which determines the size of the linear magnetoelectric effect, is related to the magnitude of the magnetic susceptibility\cite{Brown1968}.
It is posited that the ME tensor increases as the susceptibility increases. 
Magnetic frustration, which suppresses magnetic transitions, leads to an enhancement of the susceptibility at low temperatures. 
Therefore, the magnetic frustration identified in \MnTaO\ in the present study is key to understand the linear magnetoelectric effect in the system.

\section{Summary}

We conducted inelastic neutron scattering experiments on linear ME material \MnTaO , which exhibits a collinear antiferromagnetic structure. 
The spin Hamiltonian was successfully identified by LSWT calculation. 
The presence of strong frustration was demonstrated.

\section*{acknowledgements}
We are grateful to D. Kawana, T. Asami, and R. Sugiura for supporting us in the neutron scattering experiment at HRC and HER. 
The neutron experiment using HRC spectrometer at the Materials and Life Science Experimental Facility of the J-PARC was performed under a user program (Proposal No. 2021S01).
A portion of this research used resources at the High Flux Isotope Reactor, a DOE Office of Science User Facility operated by the Oak Ridge National Laboratory (ORNL).
Travel expenses for the neutron scattering experiments performed using CTAX at ORNL, USA was supported by the U.S.-Japan Cooperative Research Program on Neutron Scattering (proposal no. 2021-4). 
The neutron experiment using HER at JRR-3 was carried out by the joint research in the Institute for Solid State Physics, the University of Tokyo (Proposal No. 21403). 
H. Kikuchi was supported by Support for Pioneering Research Initiated by Next Generation (SPRING) of Japan Science and Technology Agency (JST).
S. Hasegawa was supported by the Japan Society for the Promotion of Science through the Leading Graduate Schools (MERIT) 
This project was supported by JSPS KAKENHI Grant Numbers 19KK0069 and 21H04441.


\section*{References}
\bibliographystyle{iopart-num_mod.bst}
\providecommand{\newblock}{}

\end{document}